\documentclass[12pt]{article}

\usepackage{amssymb}
\def\lddots{\mathinner{\mkern1mu\raise1pt\hbox{.}\mkern2mu
\raise4pt\hbox{.}\mkern2mu\raise7pt\vbox{\kern7pt\hbox{.}}\mkern1mu}}
\makeatletter
\def\numberbysection{\@addtoreset{equation}{section}
\def\theequation{\thesection.\arabic{equation}}}
\makeatother

\numberbysection

\newcommand{\be}{\begin{eqnarray}}
\newcommand{\ee}{\end{eqnarray}}
\newcommand{\non}{\nonumber}

\begin{document}

\begin{titlepage}
\vskip 0.4cm
\strut\hfill
\vskip 0.8cm
\begin{center}

%\begin{center}
{\LARGE On reflection algebras and twisted Yangians}

\vspace{10mm}

{\Large Anastasia Doikou\footnote{e-mail: doikou@lapp.in2p3.fr}}

\vspace{14mm}

\emph{ Laboratoire d'Annecy-le-Vieux de Physique Th{\'e}orique\\
LAPTH, CNRS, UMR 5108, Universit{\'e} de Savoie\\
B.P. 110, F-74941 Annecy-le-Vieux Cedex, France}

\end{center}

\vfill

\begin{abstract}

It is known that integrable models associated to rational $R$ matrices give rise to certain non-abelian symmetries known as Yangians. Analogously `boundary' symmetries arise when general but still integrable boundary conditions are implemented, as
originally argued by Delius, Mackay and Short from the field theory point of view, in the context of the principal chiral model on the half line. In the present study we deal with a discrete quantum mechanical system with boundaries, that is the $N$ site $gl(n)$ open quantum  spin chain.\\
In particular, the open spin chain with two distinct types of boundary condition known as soliton preserving and soliton non-preserving is considered. For both types of boundaries we present a unified framework for deriving the corresponding boundary non-local charges directly at the quantum level.  The non-local charges are simply coproduct realizations of particular
boundary quantum algebras called `boundary' or twisted Yangians, depending on the choice of boundary conditions. Finally, with the help of  linear intertwining relations between the solutions of
the reflection equation and the generators of the boundary or twisted Yangians we are able to exhibit the symmetry of the open spin chain, namely we show that a number of the boundary non-local charges are in fact conserved quantities.

\end{abstract}

\vfill
%MSC number: 81R50, 17B37
%\vfill
\rightline{LAPTH-1035/04}
%\rightline{hep-th/04}
\rightline{March 2004}
\baselineskip=16pt
\end{titlepage}
%%%%%%%%%%%%%%%%%%%%%%%%%%%%%%%%%%%%%%%%%%%%%%%%%%%%%%%%%%%%%%%%%%%%%%
%\newpage

\section{Introduction}

Symmetry breaking mechanisms have been the subject of immense interest in modern physics. A well known symmetry breaking process in the context of two dimensional integrable systems is the
implementation of general  boundaries that preserve however integrability \cite{cherednik, sklyanin}. The presence of general integrable boundaries usually reduces the original symmetry of the system giving rise to `boundary' algebraic structures known as boundary quantum groups (algebras) (see e.g. \cite{mene}--\cite{doikoun}). The boundary quantum groups are essentially
subalgebras of the usual quantum groups \cite{jimbo}--\cite{cha}, and they provide the underlying algebraic structures in the reflection equation \cite{cherednik} exactly as quantum groups do in the Yang-Baxter equation \cite{baxter}--\cite {korepin}. As is well known  the Yang--Baxter and reflection equations are collections of algebraic constraints ruling two dimensional
integrable models with boundaries. The study of such boundary symmetries for a particular class of integrable systems will be the main objective of this investigation.

The present work may be seen as the continuation of the
investigation undertaken in \cite{doikoun}, where the boundary quantum group generators for the open XXZ spin chain, were constructed  by studying the asymptotics of the open spin chain. Historically, boundary quantum group generators were obtained for the first time in the context of the sine-Gordon model in the `free fermion' point \cite{mene}, whereas in \cite{nepo}
realizations of such generators were constructed for models associated to higher rank algebras. In \cite{dema} the boundary quantum group was derived for the affine Toda field theories on the half line, and solutions of the reflection equation associated to a certain type of boundary conditions were found. Also,  boundary non--local charges were constructed classically, in the framework of the principal chiral model on the half line \cite{dema2}, for two distinct types of boundary conditions known as soliton-preserving ({\it SP}) \cite{deve}--\cite{done} and soliton non-preserving ({\it SNP}) \cite{cor}--\cite{mac}.

In this paper we focus on the quantum mechanical system that is the $N$ site $gl(n)$ open quantum spin chain, and we consider two types of boundary conditions i.e. {\it SP} and {\it SNP}. For both types of boundaries the corresponding non-local charges are constructed in a systematic way by studying the asymptotics of the
$gl(n)$ open spin chain. It turns out that the non-local charges are simply coproducts of certain boundary quantum algebras called `boundary' or twisted Yangians \cite{twist}--\cite{twist3}, depending on the choice of boundary conditions. It is worth remarking that
in \cite{dema2} linear intertwining relations involving the boundary algebra generators were used as the starting point for deriving solutions of the reflection equation for both types of boundary conditions. In the present study on the other hand we start our analysis having at our disposal c-number solutions of the reflection equation, and we simply  exploit the existence of the intertwining relations in order 
to derive the symmetry of the open spin chain (see also \cite{doikoun}).  In fact, we show explicitly that a set of the boundary non-local charges are conserved quantities, that is they commute with the transfer matrix of the open spin chain. It should be pointed out that our results rely on purely algebraic grounds, and therefore they are independent of the choice of representation.

\section{The underlying algebras}

In general, two types of spin chains exist known as closed (e.g. periodic boundary conditions) and open. To construct and study a periodic spin chain one has first to introduce the basic building block, namely the $R$ (${\cal L}$) matrix satisfying the Yang--Baxter equation \cite{baxter}--\cite{korepin}. The construction of an open spin chain on the other hand requires the
consideration of one more fundamental object called the ${\cal K}$ matrix, which satisfies another set of algebraic constraints known as the reflection equation \cite{cherednik}. The main objective of the two subsequent sections is to introduce the aforementioned
fundamental objects, and also briefly describe the corresponding algebraic framework.

\subsection{The Yang-Baxter equation}

Let us first introduce the $gl(n)$ $R$ matrix, which is a solution of the Yang-Baxter equation \cite{baxter}--\cite{korepin} \be
R_{12}(\lambda_{1}-\lambda_{2})\ R_{13}(\lambda_{1})\
R_{23}(\lambda_{2}) =R_{23}(\lambda_{2})\ R_{13}(\lambda_{1})\ R_{12}(\lambda_{1}-\lambda_{2}),\label{YBE} \ee acting on ${\mathbb V} \otimes{\mathbb V} \otimes {\mathbb V}$, and as usual $~R_{12} = R\otimes {\mathbb
I}, ~~R_{23} = {\mathbb I} \otimes R~$ and so on.
The $R$ matrix may be written in the following compact form \be R(\lambda) = {\mathbb I} +{i \over \lambda}{\cal P} \label{r0}\ee where ${\cal P}$ is the
permutation operator, acting on $({\mathbb C}^{n})^{\otimes 2}$ with \be  {\cal P}\ (a \otimes b)\ = b \otimes a ~~~~\mbox{and} ~~~~~{\cal P}^{2} ={\mathbb I}.\ee In addition, the $R$ matrix satisfies the unitarity condition, \be R(\lambda)\ \hat R(-\lambda)
\propto {\mathbb I} ~~~~\mbox{where} ~~~~~\hat R(\lambda) = {\cal P}\ R(\lambda)\ {\cal P}. \label{uni} \ee Notice that $R(\lambda) \in
\mbox{End}({\mathbb C}^{n} \otimes {\mathbb C}^{n})$, however in general one may define the object ${\cal L}(\lambda) \in \mbox{End}({\mathbb C}^{n}) \otimes {\cal Y}[\lambda^{-1}]$, where the second `space' is not represented, but it is occupied by elements of the algebra ${\cal Y}$ called the $gl(n)$ Yangian
\cite{drinf1, drinf}, and defined by the following fundamental algebraic relation, \be R_{12}(\lambda_{1} -\lambda_{2})\ {\cal L}_{13}(\lambda_{1})\ {\cal L}_{23} (\lambda_{2})= {\cal L}_{23}(\lambda_{2})\ {\cal L}_{13}(\lambda_{1})\ R_{12}(\lambda_{1} -\lambda_{2}) . \label{rtt}\ee  It is worth pointing out that the asymptotic expansion of ${\cal L}(\lambda)$ to powers of $\lambda^{-1}$ yields the generators of the Yangian, which 
satisfy exchange relations dictated by (\ref{rtt}). A more detailed analysis on the Yangian will be presented in section 3. A solution of the fundamental equation (\ref{rtt}), which we shall use hereafter, may take the following simple form \be {\cal L}(\lambda) = {\mathbb I} + {i \over \lambda} {\mathbb P} \label{l} \ee ${\mathbb P}$ is an $n \times n$ matrix with entries
${\mathbb P}_{ab} \in gl(n)$.

The Yangian (\ref{rtt}) is a Hopf algebra equipped  with a coproduct $\Delta: {\cal Y} \to {\cal Y} \otimes {\cal Y}$ \be (\mbox{id} \otimes \Delta){\cal L}(\lambda) = {\cal L}_{13}(\lambda)\ {\cal L}_{12}(\lambda), \label{co1} \ee and by treating ${\cal L}$ as an $n\times n$ matrix with entries ${\cal L}_{ab}$ being elements of ${\cal Y}$, we conclude that \be \Delta({\cal L}_{ab}(\lambda)) = \sum_{c=1}^n {\cal L}_{cb}(\lambda) \otimes {\cal L}_{ac}(\lambda), ~~~~a,~b \in \{1, \ldots, n\}. \label{cob1} \ee It will be helpful for the following to also introduce $\Delta': {\cal Y} \to {\cal Y}\otimes {\cal Y}$ obtained from $\Delta$  by permutation. In particular, let $\Pi$ be the `shift operator' $~~~\Pi:\ {\cal O}_{1} \otimes {\cal
O}_{2} \to\ {\cal O}_{2} \otimes {\cal O}_{1}~~~$  then one may write \be \Delta^{'}(x)\ = \Pi \circ \Delta(x) ,~~~x \in {\cal Y} \label{prime} \ee Also, by iteration the $l$ coproduct $\Delta^{(l)}: {\cal Y} \to {\cal Y}^{\otimes (l)}$ may be written as \be \Delta^{(l)} = (\mbox{id} \otimes \Delta^{(l-1)}) \Delta.\label{cop2} \ee By considering tensor products of ${\cal Y}$ one may construct the periodic `spin chain'.  Let us first
define the algebraic monodromy matrix $T$ as tensor product of $N$ ${\cal L}$ matrices, i.e. \be T_{0}(\lambda) = (\mbox{id}\otimes \Delta^{(N)}) {\cal L}(\lambda) ={\cal L}_{0N}(\lambda) \ldots {\cal L}_{01}(\lambda), \label{tt} \ee the monodromy matrix $T \in
\mbox{End}({\mathbb C}^{n}) \otimes {\cal Y}^{\otimes N}$ is also a solution of (\ref{rtt}). Traditionally the indices $i \in \{1, \ldots, N \}$ are associated to the `quantum spaces' and they are suppressed from the monodromy matrix (\ref{tt}), whereas the index $0$ corresponds to the so called `auxiliary space'. Finally, the transfer matrix of the periodic spin chain is
derived by simply taking the trace over the auxiliary space, \be t(\lambda) = Tr_{0}\ T_{0}(\lambda), \label{tr} \ee and it is clear that $t(\lambda) \in {\cal Y}^{\otimes N}$. It follows immediately from (\ref{rtt}) that the transfer matrix (\ref{tr}) provides a family of commuting operators \be \Big [t(\lambda),\
t(\lambda') \Big ] =0 \ee ensuring the integrability of the model. Each `quantum space' in (\ref{tt}), (\ref{tr}) is associated to a copy of ${\cal Y}$, and the corresponding sequence of $N$ copies of ${\cal Y}$ described by (\ref{tt}), (\ref{tr}) is a purely
algebraic construction. It acquires a physical meaning as a spin chain once the `quantum spaces' are mapped to finite or infinite dimensional spaces. Then the spectrum of the transfer matrix and the corresponding Bethe ansatz equations can be derived and the physically relevant quantities may be computed \cite{FT}.

\subsection{The reflection equation}

In the following two distinct types of boundary conditions {\it SP}, {\it SNP} are described using a unified framework, and the corresponding spin chains are constructed. As mentioned, in order to build the open spin chain an additional fundamental object
needs to be considered, that is the ${\cal K}^*$ matrix, which is a solution of the reflection equation \cite{cherednik} \begin{equation}
R_{12}(\lambda_{1}-\lambda_{2})\ {\cal K}^*_{1}(\lambda_{1})\ R^*_{21}(\lambda_{1}+\lambda_{2})\ {\cal K}^*_{2}(\lambda_{2})= {\cal K}^*_{2}(\lambda_{2})\ R^*_{12}(\lambda_{1}+\lambda_{2})\ {\cal K}^{*}_{1}(\lambda_{1})\ R_{21}(\lambda_{1}-\lambda_{2}) \label{re} \end{equation} acting on ${\mathbb V}\otimes {\mathbb V}$,
and as  customary $~{\cal K}^{*}_{1} = {\cal K}^{*} \otimes {\mathbb I}$, $~ {\cal K}^{*}_{2} = {\mathbb I} \otimes  {\cal K}^{*} $. We also introduce the notation, 
\be {\cal K}^{*}(\lambda) &=& {\cal K}(\lambda),\ ~~~R^{*}(\lambda) =R(\lambda) ~~~\mbox{for {\it SP} b.c.} \non\\ {\cal K}^{*}(\lambda) &=&  \bar {\cal K}(\lambda),\
~~~~R^{*}(\lambda) =\bar R(\lambda) ~~~\mbox{for {\it SNP} b.c.} \label{*} \ee and we define \be && \bar R_{12}(\lambda) = f(\lambda)\ V_{1}\ R_{12}^{t_{2}}(-\lambda-i\rho)\ V_{1}=f(\lambda)\ V_{2}^{t_{2}}\ R_{12}^{t_{1}}(-\lambda-i\rho)\ V_{2}^{t_{2}}, \non\\ && \rho = {n \over 2},~~~~~f(\lambda)={\lambda +i\rho \over \lambda} \label{rbar}  \ee  $\bar R_{21}(\lambda) = \bar R_{12} (\lambda)^{t_{1}t_{2}}$, $t_{i}$ denotes the transposition on the $i^{th}$ space. $V$ is the
`charge conjugation' being of the form \be V
&=&antidiag(1,1,...,1,1),\ n\ ~~~\mbox{odd\ and\ even}, ~~~\mbox{or} \non\\ V &=&antidiag(i,-i,...,i,-i),\ n\ ~~~\mbox{even\ only}. \label{conj} \ee The $\bar R$ matrix may be also written in a compact form as \be \bar R(\lambda) &=&{\lambda +i \rho \over \lambda} {\mathbb I} -{i \over \lambda} K \non\\ &=& {\mathbb I} +
{ i \over \lambda} \check {\cal P} \label{rb} \ee where $K$ is an one dimensional projector satisfying \be K\ {\cal P} = {\cal P}\ K= \pm K, ~~~~~K^{2} = n K, \ee and consequently $\check {\cal P}^{2} = \rho^{2} {\mathbb I}$. For the special case $n=2$, it is clear that $\check {\cal P} ={\cal P}$. Relation (\ref{re}) defines the so
called reflection algebra (boundary Yangian) \cite{sklyanin}, \cite{kulish}--\cite{avdo} or the twisted Yangian \cite{twist, twist2, twist3} depending on the action of $*$ (\ref{*}). Let us denote the boundary or twisted Yangian as ${\mathbb B}^*$, more specifically \be{\mathbb B}^*= {\mathbb B} ~~~\mbox{boundary
Yangian for {\it SP} b.c.}, ~~~~~{\mathbb B}^*= \bar {\mathbb B} ~~~\mbox{twisted Yangian for {\it SNP} b.c.} \ee It should be noted that from the physical point of view {\it SP} b.c. describe the reflection of a soliton to a soliton, while {\it SNP} b.c. describe the reflection of a soliton to an anti-soliton.

The general form of the $gl(n)$ $c$--number ${\cal K}^*$ matrix is given by \cite{cherednik} \be {\cal K}^*(\lambda) = \lambda k^*+ f^* \label{asyk} \ee  where $f^*$, $k^*$ are $n\times n$ $\lambda$ independent matrices with $~f^* =f ~\mbox{or}\ ~\bar f~$ and $~k^*=k\ ~\mbox{or}\ ~\bar k~$ for {\it SP} and {\it SNP} b.c. respectively. For {\it SP} b.c. in particular, $~f= i\xi
{\mathbb I}~$ and $~k~$ is a $n\times n$ $\lambda$ independent matrix with non-zero entries given by \cite{abad} \be k_{11}(\lambda) &=& -1, ~~~~k_{nn}(\lambda)= 1, ~~~~ k_{1n}(\lambda) =k_{n1}(\lambda) = 2\kappa\non\\
k_{jj}(\lambda)&=& 1+2c,\ ~~~j=2,\ldots,n-1,  \label{k} \ee  with $\xi$ arbitrary constant, and $c$, $\kappa$ constants satisfying $\kappa^{2}=c(c+1)$ (see also \cite{abad}), so there are two arbitrary boundary parameters $\xi$ and $\kappa$. Notice that the entries of the ${\cal K}$ matrix are simply $c$-numbers and this is the reason why (\ref{k}) is called a $c$-number solution of the
reflection equation.

Although certain solutions $\bar {\cal K}$ ({\it SNP}) of
(\ref{re}) have been derived for the trigonometric \cite{dema, gand} and rational case \cite{dema2, mac}, the situation is not completely clear yet for this type of boundary conditions.  Nevertheless, assuming the general form (\ref{asyk}) for the $c$-number $\bar {\cal K}$
matrix is sufficient for our analysis.  It is evident that the $gl(2)$ case is rather special, because $R_{12} =\bar R_{12}$ (by choosing the second $V$ in (\ref{conj})), and consequently ${\cal K} = \bar {\cal K}$.

Having at our disposal c-number solutions of the (\ref{re}) we may build the more general form of solution. To do so it is first necessary to define the following objects \be \hat {\cal L}_{12}(\lambda) = {\cal L}_{12}^{-1}(-\lambda), ~~~~~ \bar {\cal L}'(\lambda) = f(\lambda)\ V_{1}^{t_{1}}\ {\cal L}_{12}^{t_{1}}(-\lambda -i\rho)\ V_{1}^{t_{1}} \ee and \be {\cal L}^{*}(\lambda) =\hat {\cal L}(\lambda), ~~~\mbox{for {\it SP}
b.c.} ~~~~{\cal L}^{*}(\lambda) = \bar {\cal L}'(\lambda)
~~~~~\mbox{for {\it SNP} b.c.}\ee It will be instructive for the following to present explicit expressions of the ${\cal L}^*$. Although the expressions for ${\cal L}^{-1}$ and consequently $\hat {\cal L}$ are quite intricate, fortunately for our purposes it is only necessary to consider the asymptotic behavior as $\lambda \to \infty$, keeping terms up to ${1\over \lambda^2}$ \be
\hat {\cal L}_{12}(\lambda \to \infty) \propto {\mathbb I} +{i \over \lambda} {\mathbb P}_{12} -{1\over \lambda^2} {\mathbb P}_{12}^2 + {\cal O}({1\over \lambda^3}), ~~~~\mbox{and} ~~~~~\bar
{\cal L}'_{12}(\lambda) = {\mathbb I} +{i\over \lambda} \check {\mathbb P}_{12} \ee where \be \check {\mathbb P}_{12} = \rho {\mathbb I} - V_{1}^{t_{1}}\ {\mathbb P}_{12}^{t_{1}}\ V_{1}^{t_{1}}. \label{iso} \ee And more specifically, $\hat R(\lambda)$ is provided by (\ref{uni}), and $~\bar R'(\lambda) = {\cal P}\ \bar R(\lambda)\ {\cal P}$.

The more general solution of (\ref{re}) is then given by
\cite{sklyanin}: \be {\mathbb K}^*(\lambda) = {\cal
L}(\lambda-\Theta)\ ({\cal K}^{*}(\lambda)\otimes {\mathbb I})\  {\cal L}^*(\lambda
+\Theta). \label{gensol} \ee $\Theta$ some times is called inhomogeneity and henceforth for simplicity we shall consider it to be zero. The entries of ${\mathbb K}^*$ are elements of the ${\mathbb B}^{*}$ algebra defined by (\ref{re}). It is clear that the general solution (\ref{gensol}) allows the  expansion in
powers of $\lambda^{-1}$ as we shall see in subsequent sections, providing the generators of the boundary Yangian (for ${\it SP}$) or the twisted Yangian (for ${\it SNP}$), which obey commutation relations dictated by the defining algebraic relations (\ref{re}).
The algebra ${\mathbb B}^*$ is also endowed with a coproduct inherited essentially from the Yangian. In particular, let us first derive the coproduct for ${\cal L}^*$, i.e. \be (\mbox{id} \otimes \Delta)  {\cal L}^*(\lambda) =  {\cal L}^*_{12}(\lambda)\
{\cal L}^*_{13}(\lambda) \rightarrow \Delta({\cal
L}^*_{ab}(\lambda)) = \sum_{c=1}^n {\cal L}^*_{ac}(\lambda) \otimes  {\cal L}^*_{cb}(\lambda) ~~~~a,~b\in\{1, \ldots, n\}. \label{cob2} \ee Then it is clear from (\ref{cob1}), (\ref{cob2}) that the elements of ${\mathbb B}^{*}$ form coproducts $\Delta:
{\mathbb B}^* \to {\mathbb B}^* \otimes {\cal Y}$, such that (see also \cite{dema, dema2}) \be \Delta({\mathbb K}^*_{ab}(\lambda)) = \sum_{k,l=1}^n{\mathbb K}^*_{kl}(\lambda) \otimes {\cal L}_{ak}(\lambda) {\cal L}^*_{lb}(\lambda)~~~~a,~b\in\{1, \ldots, n\}. \label{coc} \ee Our final aim of course is to build the
corresponding quantum system that is the open quantum spin chain. For this purpose we shall need tensor product realizations of the general solution (\ref{gensol}). We define \be T^*_{0} (\lambda) = (\mbox{id} \otimes \Delta^{(N)}) {\cal L}^*(\lambda) = {\cal L}_{01}^*(\lambda)\ldots  {\cal L}_{0N}^* (\lambda)
\label{tt'} \ee then the  general tensor type solution of the (\ref{re}) takes the form \be {\cal T}_{0}^* (\lambda) = T_{0}(\lambda)\ {\cal K}_{0}^*(\lambda)\ T_{0}^*(\lambda), \label{skl} \ee with entries being
clearly coproducts of the ${\mathbb B}^*$ algebra, namely \be {\cal T}^*_{ab}(\lambda) =\Delta^{(N)}({\mathbb K}_{ab}(\lambda)) \label{bbco} \ee recall
that $\Delta^{(N)}$ is defined via (\ref{cop2}), and
$\Delta({\mathbb K}_{ab})$ is given in (\ref{coc}). Finally, we introduce the transfer matrix of the open spin chain \cite{sklyanin}, which may be written as \be t^{*}(\lambda) = Tr_{0}\  \Big \{ {\cal K}_{0}^{(+)*}(\lambda)\ {\cal T}^{*}_{0}(\lambda) \Big \}, \label{transfer} \ee  
${\cal K}^{(+)*}(\lambda) ={\cal K}^{(l)*}(-\lambda-i\rho)$, where ${\cal K}^{(l)*} (\lambda)$ is also a solution of (\ref{re}), but here for simplicity is considered tobe ${\mathbb I}$. The $*$ notation for the aforementioned objects is applied as follows \be && T^*(\lambda) = \hat T(\lambda),
~~~~{\cal T}^*(\lambda) ={\cal
T}(\lambda), ~~~~t^*(\lambda)=t(\lambda),~ ~~~~\mbox{for {\it SP} b.c.}, \non\\
&&  T^*(\lambda) = \bar T(\lambda), ~~~~{\cal T}^* (\lambda) =\bar{\cal T}(\lambda), ~~~~t^*(\lambda)=\bar t(\lambda),~~~ \mbox{for {\it SNP} b.c.} \ee It can be shown \cite{sklyanin, doikou}, using
the fact that ${\cal T}^*$ is a solution of the reflection equation (\ref{re}), that the transfer matrices (\ref{transfer})
provide families of commuting operators i.e., \be \Big
[t^{*}(\lambda),\ t^{*}(\lambda')\Big ] =0. \label{com} \ee The latter commutation relations (\ref{com}) ensure the integrability of the relevant models.

\section{More on Yangians}

It is instructive for what follows to recall in more detail the basic definitions associated to Yangians \cite{drinf1,  bernard} (for a review on Yangians see e.g. \cite{molev}). The $gl(n)$ Yangian ${\cal Y}$, is a non abelian algebra ---a quantum group \cite{drinf}--\cite{cha}--- with generators ${\cal Q}_{ab}^{(p)}$
and defining relations  given below \be &&\Big [ {\cal
Q}_{ab}^{(0)},\ {\cal Q}_{cd}^{(0)} \Big ] =i\delta_{cb}{\cal Q}^{(0)}_{ad} - i\delta_{ad}{\cal Q}^{(0)}_{cb} \non\\ && \Big [ {\cal Q}_{ab}^{(0)},\ {\cal Q}_{cd}^{(1)} \Big ] =i\delta_{cb}
{\cal Q}^{(1)}_{ad} - i\delta_{ad}{\cal Q}^{(1)}_{cb} \non\\  && \Big [ {\cal Q}_{ab}^{(1)},\ {\cal Q}_{cd}^{(1)} \Big ]
=i\delta_{cb}{\cal Q}^{(2)}_{ad} -i \delta_{ad}{\cal Q}^{(2)}_{cb}
+{i h^2\over 4}{\cal Q}_{ad}^{(0)}(\sum_{e} {\cal
Q}_{ce}^{(0)}{\cal Q}_{eb}^{(0)})- {i h^2\over 4}(\sum_{e}{\cal Q}_{ae}^{(0)}{\cal Q}_{ed}^{(0)}){\cal Q}_{cb}^{(0)} \non\\ && a, b \in \{ 1,\ldots, n \} \ee and also relations \be \Big [ {\cal Q}_{ab}^{(0)},\ \Big [ {\cal Q}_{cd}^{(1)},\ {\cal Q}_{ef}^{(1)} \Big ] \Big ]&-& \Big [ {\cal Q}_{ab}^{(1)},\ \Big [ {\cal Q}_{cd}^{(0)},\ {\cal Q}_{ef}^{(1)} \Big ] \Big ] \non\\ =
{h^2\over 4}\sum_{p,q} \Big ( \Big [ {\cal Q}_{ab}^{(0)},\ \Big [ {\cal Q}_{cp}^{(0)}{\cal Q}_{pd}^{(0)},\ {\cal Q}_{eq}^{(0)}{\cal
Q}_{qf}^{(0)} \Big ] \Big ] &-&\Big [ {\cal Q}_{ap}^{(0)}{\cal Q}_{pb}^{(0)},\ \Big [ {\cal Q}_{cd}^{(0)},\ {\cal
Q}_{eq}^{(0)}{\cal Q}_{qf}^{(0)} \Big ] \Big ]  \Big ). \ee As already mentioned (\ref{cob1}) the Yangian is endowed with a coproduct $\Delta: {\cal Y} \to {\cal Y} \otimes {\cal Y}$. In particular, the coproducts of the generators ${\cal Q}_{ab}^{(p)}$ may be written as \be \Delta({\cal Q}_{ab}^{(0)})&=& {\cal
Q}_{ab}^{(0)}\otimes {\mathbb I} + {\mathbb I}\otimes {\cal Q}_{ab}^{(0)} \non\\  \Delta({\cal Q}_{ab}^{(1)})&=& {\cal Q}_{ab}^{(1)} \otimes {\mathbb I} + {\mathbb I}\otimes {\cal Q}_{ab}^{(1)} +{h \over 2} \sum _{d=1}^n({\cal Q}_{ad}^{(0)}\otimes {\cal Q}_{db}^{(0)}- {\cal
Q}_{db}^{(0)}\otimes {\cal Q}_{ad}^{(0)}), \label{cop} \ee also for $\Delta'$ similar expressions may be deduced. In fact the only difference between expressions $\Delta$ and $\Delta'$ is a minus sign in front of $h$ in the coproduct of ${\cal Q}_{ab}^{(1)}$. Using (\ref{cop2}) we can get explicit expression for the $l$
coproducts \be \Delta^{(l)}({\cal Q}_{ab}^{(0)}) &=& 
\sum_{i=1}^{l}({\cal Q}_{ab}^{(0)})_{i} \non\\
\Delta^{(l)}({\cal Q}_{ab}^{(1)})&=& \sum_{i=1}^{l}({\cal
Q}_{ab}^{(1)})_{i}+{h \over 2} \sum_{j>i=1}^{l} \sum _{d=1}^n \Big (({\cal Q}_{ad}^{(0)})_{i}\otimes ({\cal Q}_{db}^{(0)})_{j}- ({\cal Q}_{db}^{(0)})_{i}\otimes ({\cal Q}_{ad}^{(0)})_{j} \Big), \label{ncop}\ee where the indices $i,\ j$ denote the site in the $l$ coproduct sequence. We may also define $\Delta^{'(l)}: {\cal
Y} \to {\cal Y}^{\otimes (l)}$ as \be \Delta^{'(l)} = (\mbox{id} \otimes \Delta^{(l-1)}) \Delta'. \label{cop22} \ee Note that there also exists the opposite coproduct $~\Delta^{op(l)} = (\mbox{id}
\otimes \Delta^{op(l-1)}) \Delta^{op}~$ with $~\Delta^{op}= \Delta'$. For generic values of $l$, $~\Delta^{op(l)}\neq \Delta^{'(l)}~$ and they only coincide for $l=2$. Expressions similar to (\ref{ncop}) may be derived for $\Delta^{'(l)}$, but we omit them here for brevity.

\subsection*{Realizations of Yangian generators}

As is known the asymptotic behaviour of the monodromy matrix $T$
(\ref{tt}) as $\lambda \to \infty$ provides tensor product representations of ${\cal Y}$. Let us briefly review how this process works. Recall that the operators ${\cal L}$ and $T$ are treated as $n \times n$ matrices with entries being elements of ${\cal Y}$, ${\cal Y}^{\otimes N}$ respectively. The monodromy matrix 
$T$ as $\lambda \to \infty$ may be written as
(for simplicity we suppress the `auxiliary' space index $0$ from $T$ in the following) \be T(\lambda \to \infty) \propto {\mathbb I} + \sum_{m=0}^{\infty} \lambda^{-m-1}\ t^{(m)}. \label{asy} \ee
Exchange relations among the charges $t_{ab}^{(m)}$ (the entries of $t^{(m)}$) may be derived by virtue of the fundamental algebraic relation (\ref{rtt}), as $\lambda_{i} \to \infty$. To extract the Yangian generators we study the asymptotic expansion (\ref{asy}) keeping higher orders in the ${1 \over \lambda}$ expansion. Recalling the form of ${\cal L}$ (\ref{l}) and $T$ (\ref{tt}) we conclude that \be T(\lambda \to \infty) \propto {\mathbb I} +{i\over \lambda}\sum_{i=1}^{N} {\mathbb P}_{0i} -{1\over \lambda^{2}} \sum_{i > j=1}^{N} {\mathbb P}_{0i}\ {\mathbb P}_{0j} +{\cal O}({1\over \lambda ^{3}}). \label{asyt} \ee Now consider the quantities below written as combinations of $t^{(p)},~~p \in \{0,\ 1 \}$,  \be Q^{(0)} = t^{(0)}, ~~~~~Q^{(1)} = t^{(1)}
-{1\over 2} Q^{(0)}\ Q^{(0)} \ee where the from of $t^{(p)}$ is defined by (\ref{asy}), (\ref{asyt}). Then $Q^{(p)}$ may be written  as combinations of the operators ${\mathbb P}_{0i}$, each acting on ${\mathbb C}^{n} \otimes {\cal Y}$, namely (from now on we consider $h=-1$ in (\ref{cop})) \be Q^{(0)} = i\sum_{i=1}^{N} {\mathbb P}_{0i},~~~~~Q^{(1)} = {1\over 2}\sum_{i=1}^{N}{\mathbb P}_{0i}^2 +{1\over 2}\sum_{j>i=1}^{N}
({\mathbb P}_{0i}\ {\mathbb P}_{0j} - {\mathbb P}_{0j}\ {\mathbb P}_{0i}). \label{qb} \ee Note that for simplicity both quantum and auxiliary indices in $Q^{(p)}$ (\ref{qb}) are omitted. The entries of the matrices $Q^{(p)}$ are the non-local charges $Q_{ab}^{(p)} \in {\cal Y}^{\otimes N}$ being coproduct realizations of the Yangian, i.e. \be Q_{ab}^{(p)} = \Delta^{(N)}({\cal Q}_{ab}^{(p)}) ~~~~p \in \{ 0,\
1\}, ~~~~a,b \in \{1, \ldots, n\}. \label{coco}\ee The charges $Q_{ab}^{(0)}$ in particular, are coproducts of the generators of the $gl(n)$ Lie algebra. It is also apparent from (\ref{coco}) that for $N=1$, $Q_{ab}^{(p)} \to {\cal Q}_{ab}^{(p)}$.

\section{Boundary and twisted Yangian generators}

After the brief review on Yangians we are in the position to deal with realizations of boundary or twisted Yangians. As already mentioned two different types of boundary conditions, the {\it SP} and the {\it SNP} \cite{cor}--\cite{mac}, \cite{our2} will be
investigated using a unified framework. The {\it SNP} boundary conditions were studied for the first time in the context of integrable lattice models 
in \cite{doikou}, whereas a generalized description of these boundaries is presented in \cite{our2}.

As we have seen in section 2.2 the entries of ${\cal T}^*$ are coproducts of the ${\mathbb B}^*$ algebra.  The main objective now is to obtain, as in the case of Yangians,  the exact form of tensor products of the generators of ${\mathbb B}^*$ via the asymptotic expansion of ${\cal T}^*$. Again ${\cal L}$(${\cal L}^*$), $T$($T^*$ ) and consequently ${\cal T}^*$ (\ref{transfer}) are treated as $n \times n$ matrices with
entries being elements of ${\cal Y}$, ${\cal Y}^{\otimes N}$ respectively. Recall also that ${\cal K}^*$ (\ref{k}) is a $n \times n$ matrix with $c$-number entries. The expansion of ${\cal T}^{*}$ (\ref{transfer}) as $\lambda \to \infty$ reads (again for simplicity we suppress the `auxiliary space' index $0$ from $T$, $T^{*}$ and ${\cal T}^{*}$ in the following) \be {\cal T}^{*}(\lambda \to \infty) \propto {\mathbb I} +
\sum_{m=0}^{\infty}\lambda ^{-m-1}\ \tilde t^{(m)}, \label{asy2} \ee while exchange relations among the charges $\tilde t_{ab}^{(m)}$ may be now found by virtue of the algebraic relations (\ref{re}).

As in the bulk case to extract the boundary or twisted Yangian generators we shall keep higher order terms in the expansion (\ref{asy2}). We need first the asymptotic behaviour of the matrices $T$, $T^*$ as well as ${\cal K}^{*}$. The expansion of $T$ is given by (\ref{asyt}), while $T^*$ reads as $\lambda \to \infty$ \be T^{*}(\lambda \to \infty) \propto {\mathbb I}
+{i\over \lambda}\sum_{i=1}^{N} {\mathbb  P}^{*}_{0i} -{1\over \lambda^{2}} (\sum_{i<j=1}^{N} {\mathbb P}^{*}_{0i} \ {\mathbb P}^{*}_{0j}+ Y^*) +{\cal O}({1\over \lambda ^{3}}). \label{asyt1} \ee where \be &&{\mathbb P}^{*} = {\mathbb P} ~~~\mbox{for {\it SP} b.c.} ,~~~~~{\mathbb P}^{*} = \check {\mathbb P} ~~\mbox{for
{\it SNP} b.c}, ~~~\mbox{and} \non\\ &&Y^* = \sum_{i=1}^N {\mathbb P}^2_{0i} ~~~ \mbox{for {\it SP} b.c.}, ~~~~~Y^* =0 ~~~\mbox{for {\it SNP} b.c.} \ee  Before we continue with the asymptotics of ${\cal T}^*$ it is necessary for our purposes to derive the charges from the expansion of $\bar T$ as $\lambda \to
\infty$ (the expansion of $\hat T$ yields the same charges as in (\ref{qb})). Note that $\bar T$ also satisfies the defining relation of the Yangian (\ref{rtt}), therefore the corresponding charges are expected to be coproducts of the Yangian generators.
In fact,  the asymptotic expansion of $\bar T$  (up to ${1\over \lambda^2}$) provides the following operators, \be \check Q^{(0)} = i\sum_{i=1}^{N} \check {\mathbb P}_{0i}, ~~~~~~\check Q^{(1)} = -{1\over 2}\sum_{i=1}^{N}\check {\mathbb P}_{0i}^2 +{1\over 2}
\sum_{j>i=1}^{N} (\check {\mathbb P}_{0i}\ \check {\mathbb P}_{0j} - \check {\mathbb P}_{0j}\ \check {\mathbb P}_{0i}). \label{qbb} \ee The
entries of the matrices (\ref{qbb}) may be indeed written as coproducts of an alternative set of generators of the Yangian $\check {\cal Q}_{ab}^{(p)}$, \be \check Q_{ab}^{(p)} = \Delta^{(N)}(\check {\cal Q}_{ab}^{(p)}), ~~~p \in \{ 0,\  1\},
~~~a,\ b \in \{1, \ldots, n \}, \ee where  $\check {\cal
Q}_{ab}^{(p)}$ are the entries of the matrices derived in
(\ref{qbb}) for $N=1$. The generators $\check {\cal
Q}_{ab}^{(p)}$are isomorphic to ${\cal Q}_{ab}^{(p)}$, and their exact correspondence may be found by exploiting the relation between ${\mathbb P}$ and $\check {\mathbb P}$ given by (\ref{iso}).

Recall also that the ${\cal K}^{*}$ matrix is given by the general form (\ref{asyk}) for any solution of the reflection equation (\ref{re}), associated to the $gl(n)$ $R$ matrix. Having derived
the expansions of $T$, $T^{*}$ and ${\cal K}^{*}$ we may now come to the asymptotic behaviour of ${\cal T}^{*}$ as $\lambda \to \infty$, (we keep here up to ${1\over \lambda^{2}}$ terms) i.e.\be &&{\cal T}^{*}(\lambda \to \infty) \propto k^{*}+{1\over
\lambda}(f^{*}+ ik^{*}\sum_{i=1}^{N} {\mathbb P}^{*}_{0i}
+i\sum_{i=1}^{N}{\mathbb P}_{0i}k^{*})+{1\over \lambda^{2}} (-k^{*}\sum_{i<j=1}^{N}{\mathbb P}^{*}_{0i}{\mathbb
P}^{*}_{0j}\non\\ &&-\sum_{i>j=1}^{N}{\mathbb P}_{0i} {\mathbb P}_{0j}k^{*} -\sum_{i,j=1}^{N}{\mathbb P}_{0i}k^{*}{\mathbb P}^{*}_{0j}+if^{*}\sum_{i=1}^{N} {\mathbb P}^{*}_{0i}+i\sum_{i=1}^{N}{\mathbb P}_{0i}f^{*} - k^* Y^*) +{\cal O}({1\over \lambda^{3}}), \label{asyt2} \ee which provides the form of $\tilde t^{(p)}, ~~p\in \{ 0,\ 1\}$ (\ref{asy2}). Consider now the following combinations of $\tilde t^{(p)}$ \be \tilde Q^{(0)} = \tilde t^{(0)}-f^*, ~~~~~\tilde Q^{(1)}= \tilde t^{(1)} -{1\over
2} \tilde Q^{(0)} (k^*)^{-1} \tilde Q^{(0)}, \label{trans} \ee where for the general $gl(n)$ solution (\ref{k}) $k^{-1} = {1\over 1+ 4\kappa^{2}}k$, and $\bar k$ ({\it SNP} b.c.) is also invertible (see e.g \cite{gand}). Let us also introduce the following objects \be {\cal D} =( {\mathbb P}k^* - k^*{\mathbb P}^*), ~~~~~~{\cal S} =(k^{*} {\mathbb P}^* + {\mathbb P}k^*),
\label{sd} \ee  \be Q^{*(p)}&=&Q^{(p)} ~~~\mbox{for {\it SP} b.c.}, ~~~~~Q^{*(p)} =\check Q^{(p)} ~~~\mbox{for {\it SNP} b.c.} \label{statm'} \ee
%%%%In fact $\bar {\cal P} = {\cal P}_{\bar b\ \bar a}$.
Then according to (\ref{qb}), (\ref{qbb}), (\ref{trans}),
(\ref{sd}) the matrices $\tilde Q^{(p)}$ may be written as
%%%$Nk$ for SP and ${N\over 2} (\rho^{2}+1)\bar k$ for SNP.
\be \tilde Q^{(0)} &=& i\sum_{i=1}^{N}{\cal S}_{0i}, \non\\ \tilde Q^{(1)} &=& f^* Q^{*(0)} +Q^{(0)}f^* +{1\over 2} \sum_{i<j=1}^{N}({\cal S}_{0i}(k^*)^{-1}{\cal D}_{0j} -{\cal D}_{0j} (k^*) ^{-1} {\cal S}_{0i})- k^*  Y^* \non\\ &+& {1\over 2} k^* \sum_{i=1}^{N} {\mathbb P}^{*2}_{0i}+{1\over 2} \sum_{i=1}^{N} {\mathbb P}^{2}_{0i} k^* +{1\over 2}\sum_{i=1}^{N}(k^*{\mathbb P}^*_{0i}(k^*)^{-1} {\mathbb P}_{0i}k^* - {\mathbb P}_{0i} k^*{\mathbb P}^*_{0i}) . \label{expr}  \ee  The corresponding entries $~\tilde Q_{ab}^{(p)} ~~ a,b \in \{1,\ldots, n\}~$ are the boundary non-local charges. Notice that the two
last terms in (\ref{expr}) vanish for the special case of ${\it SP}$  where $k={\mathbb I}$. ${\cal K}^*={\mathbb I}$ ($k^*={\mathbb I}$, $f^* =0$) is a valid solution  of (\ref{re}) for both boundary conditions. We should point out that for ${\cal K}={\mathbb I}$ ({\it SP} b.c.) $\tilde Q_{ab}^{(1)}= 0$
(\ref{expr}), so the only charges that survive are the  $\tilde Q_{ab}^{(0)}\propto Q_{ab}^{(0)}$. There exist of course higher non-local charges that may be obtained by keeping higher order terms in the asymptotic expansion (\ref{asyt2}). This is a significant investigation, which will be undertaken however in a forthcoming work.

The non--local charges $\tilde Q_{ab}^{(p)}$ (\ref{expr}) may be written as combinations of the Yangian coproducts $Q_{ab}^{(p)}$, $\check Q_{ab}^{(p)}$ (\ref{qb}), (\ref{qbb}) \be \tilde Q_{ab}^{(0)} &=& k^*_{ac} Q_{cb}^{*(0)}+Q_{ac}^{(0)}k^*_{cb},
\non\\ \tilde Q_{ab}^{(1)} &=&-k^*_{ac} Q_{cb}^{*(1)}+Q_{ac}^{(1)} k^*_{cb} + f^*_{ac} Q_{cb}^{*(0)} + Q_{ac}^{(0)} f^*_{cb} \non\\ &-&{1\over
2}(k^*_{ac}Q^{*(0)}_{cd}(k^*)^{-1}_{de}Q^{(0)}_{ef}k^*_{fb} -Q^{(0)} _{ac} k^*_{cd} Q^{*(0)}_{db}), ~~~~a,\ b \in \{ 1, \ldots, n \}. \label{tilde} \ee Note that the summation over repeated indices is omitted from now on. The quantities $k^*_{ab}$, $f^*_{ab}$ are $c$ numbers (the entries of the matrices $k^*$, $f^*$), $Q_{ab}^{(p)}$ are given by (\ref{qb}) and $Q_{ab}^{*(p)}$ by (\ref{statm'}). The derivation of the boundary non--local charges (\ref{tilde}) is one of the main results of this study. It is worth emphasizing that the  non-local
charges (\ref{tilde}) were derived independently of the choice of representation.

\section{The symmetry}

Our ultimate goal is to study the symmetry 
of the open spin chain, namely to derive conserved 
quantities commuting with the open transfer matrix. This may be achieved by using linear intertwining relations between representations of the boundary Yangian generators and the solutions of the reflection equation.

In the previous section we  derived the boundary
non-local charges (\ref{tilde}) as coproducts of the  boundary or twisted Yangian generators. It is evident that from the expressions (\ref{tilde}) for $N=1$, one may write down the corresponding abstract generators of ${\mathbb B}^*$, \be \tilde
{\cal Q}_{ab}^{(0)} &=& k^*_{ac} {\cal Q}_{cb}^{*(0)}+{\cal Q}_{ac}^{(0)}k^*_{cb}, \non\\ \tilde {\cal Q}_{ab}^{(1)} &=&-k^*_{ac} {\cal Q}_{cb}^{*(1)}+
{\cal Q}_{ac}^{(1)} k^*_{cb} + f^*_{ac} {\cal Q}_{cb}^{*(0)}+ {\cal Q}_{ac}^{(0)} f^*_{cb} \non\\
&-&{1\over 2}(k^*_{ac}{\cal Q}^{*(0)}_{cd}(k^*)^{-1}_{de}{\cal Q}^{(0)}_{ef} k^*_{fb} -{\cal Q}^{(0)}_{ac}k^*_{cd}{\cal Q}^{*(0)}_{db}),~~~~a,\ b \in \{ 1, \ldots, n \}. \label{gene} \ee
The quantities $\tilde {\cal Q}_{ab}^{(0)}$ are simply linear combinations of the generators of the $gl(n)$ Lie algebra. For {\it SNP} b.c. in particular the `conjugate' generators $\check {\cal Q}_{ab}^{(0)}$ are combined together with ${\cal Q}_{ab}^{(0)}$ in the expression (\ref{gene}) for $\tilde {\cal
Q}_{ab}^{(0)}$, a fact that implies the `folding' of the $gl(n)$ algebra for $k^*={\mathbb I}$ (for more details on this subject see e.g. \cite{doikou, our2}).

As an immediate consequence of (\ref{coco}), (\ref{tilde}) and (\ref{gene}) the boundary non-local charges may be written in a more compact form as \be \tilde Q_{ab}^{(p)} = \Delta^{(N)}( \tilde {\cal Q}_{ab}^{(p)}), ~~~~p  \in \{ 0,\ 1\}. \label{coco2}
\ee The main advantage when deriving generators of ${\mathbb B}^*$ via the asymptotics of the spin chain is that one directly obtains the explicit form of the coproducts of ${\mathbb B}^*$ generators(\ref{tilde}), (\ref{coco}). Bearing in mind the coproducts  of
the Yangian generators (\ref{cop}) and also equations
(\ref{tilde}), (\ref{coco}) for $N=2$ one may derive the following more convenient expressions for $\Delta: {\mathbb B}^* \to {\mathbb B}^* \otimes {\cal Y}$ \be \Delta(\tilde {\cal Q}_{ab}^{(0)})&=& {\mathbb I} \otimes \tilde {\cal Q}_{ab}^{(0)} + \tilde {\cal Q}_{ab}^{(0)} \otimes {\mathbb I}, \non\\
\Delta(\tilde {\cal Q}_{ab}^{(1)})&=& {\mathbb I} \otimes \tilde {\cal Q}_{ab}^{(1)} + \tilde {\cal Q}_{ab}^{(1)} \otimes {\mathbb I} -{ (k^*)^{-1}_{cd} \over 2} (\tilde {\cal Q}_{ac}^{(0)} \otimes
\tilde {\cal Q}_{db}^{(0)-} - \tilde {\cal Q}_{db}^{(0)} \otimes \tilde {\cal Q}_{ac}^{(0)-}) \label{co} \ee where \be \tilde {\cal Q}_{ab}^{(0)-}= {\cal Q}_{ac}^{(0)}k^*_{cb}- k^*_{ac} {\cal Q}_{cb}^{*(0)}.  \label{co-} \ee Similarly, with the help of
(\ref{prime}) one obtains $\Delta': {\mathbb B}^* \to {\cal Y} \otimes {\mathbb B}^*$ \be \Delta^{'}(\tilde {\cal Q}_{ab}^{(1)})&=& {\mathbb I} \otimes \tilde  {\cal Q}_{ab}^{(1)} + \tilde {\cal Q}_{ab}^{(1)} \otimes {\mathbb I} +{ (k^*)^{-1}_{cd} \over 2} (\tilde {\cal Q}_{ac}^{(0)-} \otimes \tilde {\cal Q}_{db}^{(0)} - \tilde {\cal Q}_{db}^{(0)-} \otimes \tilde {\cal Q}_{ac}^{(0)}), \label{co'} \ee and the $l$ coproducts are deduced in a straightforward manner via (\ref{cop2}), (\ref{cop22}). Note that expressions similar to (\ref{gene}), (\ref{co}), (\ref{co'}), but not exactly the same, were also derived in \cite{dema2} from a field theoretical point of view. 

Consider now the evaluation representation $\pi_{\lambda}: {\cal Y} \to \mbox{End}({\mathbb C}^n) $ such that \be \pi_{\lambda}({\cal Q}_{ab}^{*(1)})=i\lambda {\cal P}^*_{ab},\ ~~~~\pi_{\lambda}( {\cal Q}_{ab}^{*(0)})= i{\cal P}^*_{ab}, ~~~~a,b \in
\{ 1, \ldots, n \}. \label{eval} \ee ${\cal P}^{*}$ is an $n \times n$ matrix, with entries ${\cal P}_{ab}^{*}$ being operators which act on $\mathbb C^{n}$. The generators (\ref{gene}) are then expressed in terms of the operators ${\cal P}_{ab}^{*}$ as \be
\pi_{\lambda}(\tilde {\cal Q}_{ab}^{(0)})&=& i k^*_{ac} {\cal P}^*_{cb}+i{\cal P}_{ac} k^*_{cb},\non\\
\pi_{\lambda}(\tilde {\cal Q}_{ab}^{(1)}) &=&-i \lambda k^*_{ac} {\cal P}^*_{cb}+ i\lambda {\cal P}_{ac}k^*_{cb} +i f^*_{ac} {\cal P}^*_{cb} +i{\cal P}_{ac}f^*_{cb} \non\\ &+&{1\over 2}(k^*_{ac}{\cal P}^*_{cd}(k^*) ^{-1}_{de}{\cal P}_{ef}k^*_{fb}-{\cal P}_{ac}k^*_{cd}{\cal P}^*_{db}). \label{tilde2} \ee We
shall henceforth restrict our attention to {\it SP} boundary conditions only, and we shall derive certain
intertwining relations between the charges (\ref{tilde2}) and the ${\cal K}$ matrix (\ref{k}). More specifically, it may be directly deduced from the reflection equation (\ref{re}) that all the elements of the algebra ${\mathbb B}$ `commute' with the ${\cal K}$ matrix (see also \cite{dema, dema2}). 
Indeed, by acting with the evaluation representation on the second space of (\ref{gensol}) we obtain \be (\mbox{id} \otimes \pi_{\pm \lambda}){\mathbb K}(\lambda') = R(\lambda' \mp \lambda)\ ({\cal K}(\lambda')
\otimes {\mathbb I})\ \hat R(\lambda \pm\lambda'). \ee Now recalling the reflection equation (\ref{re}) and because of the form of the above expressions it is straightforward to show that \be (\mbox{id} \otimes \pi_{\lambda}){\mathbb K}(\lambda')\ ({\mathbb I} \otimes {\cal K}(\lambda))= ({\mathbb I} \otimes {\cal K}(\lambda))\  (\mbox{id} \otimes \pi_{-\lambda})
{\mathbb K}(\lambda'). \ee As a consequence the entries of ${\mathbb K}$ in the evaluation representation `commute' with the c-number ${\cal K}$ matrix (\ref{k}) \be \pi_{\lambda}({\mathbb K}_{ab}(\lambda'))\ {\cal K}(\lambda) = {\cal K}(\lambda)\ \pi_{-\lambda}({\mathbb K}_{ab}(\lambda')), ~~~a,\ b \in \{1, \ldots,n \}. \label{bcomm}\ee In addition, as we have seen from the analysis of the previous section, the elements ${\mathbb K}_{ab}(\lambda' \to \infty)$ provide essentially the generators (\ref{gene}), and therefore we conclude that \be \pi_{\lambda}(\tilde {\cal Q}_{ab}^{(p)})\ {\cal K}(\lambda)= {\cal K}(\lambda)\ \pi_{-\lambda} 
(\tilde {\cal Q}_{ab}^{(p)}). \label{inter} \ee Note that we also verified by inspection, taking into account the form of the $gl(n)$ ${\cal K}$ matrices (\ref{asyk}), (\ref{k}) and (\ref{tilde2}), that the latter relations (\ref{inter}) are indeed satisfied.
 
Equations (\ref{inter}) are the boundary analogues of the bulk intertwining relations for the ${\cal L}$ matrix, i.e. \be & &(\pi_{\lambda}\otimes \mbox{id})\Delta'({\cal Q}_{ab}^{(p)})\
{\cal L}(\lambda) = {\cal L}(\lambda)\ (\pi_{\lambda}\otimes \mbox{id})\Delta({\cal Q}_{ab}^{(p)}),  \non\\
&&(\pi_{-\lambda}\otimes \mbox{id})\Delta({\cal Q}_{ab}^{(p)})\ \hat {\cal L}(\lambda) = \hat {\cal L}(\lambda)\
(\pi_{-\lambda}\otimes \mbox{id}) \Delta'({\cal Q}_{ab}^{(p)}). \label{intb} \ee  Relations of the form (\ref{inter}) should also hold for solutions $\bar {\cal K}$ of (\ref{re}) for the general $gl(n)$ case, which however merits further study and it will be the subject of a forthcoming work. It is worth remarking that in \cite{dema, dema2} intertwining relations such as in (\ref{inter}) were used as a starting point for deriving solutions of the reflection equation. 

Expressions of the type (\ref{inter}) may be obtained
for ${\cal T}$ (\ref{skl}) as well (see also
\cite{doikoun}). To derive the generalized intertwining relations for the ${\cal T}$ matrix we first need to show relations similar to (\ref{intb}) for the monodromy matrices $T$ and $\hat T$. Indeed, it immediately follows by induction using (\ref{intb}) and the definitions (\ref{tt}), (\ref{tt'}) that \be && (\pi {\lambda} \otimes \mbox{id}^{\otimes N})\Delta^{'(N+1)}( {\cal Q}_{ab}^{(p)})\ T(\lambda) = T(\lambda)\ (\pi_{\lambda} \otimes \mbox{id}^{\otimes
N})\Delta^{(N+1)}(  {\cal Q}_{ab}^{(p)}) \non\\ &&(\pi_{-\lambda} \otimes \mbox{id}^{\otimes N})\Delta^{(N+1)}({\cal Q}_{ab}^{(p)})\ \hat T(\lambda) = \hat
T(\lambda)\ (\pi_{-\lambda} \otimes \mbox{id}^{\otimes
N})\Delta^{'(N+1)}( {\cal Q}_{ab}^{(p)}), ~~~p \in \{0,\ 1 \}. \label{intbt} \ee It should be stressed that the latter relations provide also an effective means for studying the symmetry of the periodic $gl(n)$ spin chain. From (\ref{inter}) and because of the form of the coproducts (\ref{co}) we conclude also that \be
(\pi_{\lambda} \otimes \mbox{id}^{\otimes N})\Delta^{(N+1)}(\tilde {\cal Q}_{ab}^{(p)})\ {\cal K}(\lambda) = {\cal K}(\lambda)\ (\pi_{-\lambda} \otimes \mbox{id}^{\otimes N})\Delta^{(N+1)}(\tilde {\cal Q}_{ab}^{(p)}). \label{interbb}\ee Recalling that the generators of ${\mathbb B}$ (\ref{gene}) are written exclusively in terms of ${\cal Q}_{ab}^{(p)}$ we conclude that the
intertwining relations (\ref{intbt}) hold also for  $\tilde {\cal Q}^{(p)}_{ab}$. Then taking into account relations (\ref{intbt}) for  $\tilde {\cal Q}_{ab}^{(p)}$, (\ref{interbb}), and also (\ref{skl}), we may derive the following relations \be (\pi_{\lambda} \otimes \mbox{id}^{\otimes N})\Delta^{'(N+1)}(\tilde {\cal Q}_{ab}^{(p)})\ {\cal T}(\lambda) = {\cal T}(\lambda)\ 
(\pi_{-\lambda} \otimes \mbox{id} {\otimes
N})\Delta^{'(N+1)}(\tilde {\cal Q}_{ab}^{(p)}), 
\label{interg} \ee which hold for the general $gl(n)$ case. The derivation of the representations (\ref{tilde2}) and the intertwining relations (\ref{inter}), (\ref{interg}) are also among the main results of this article. The latter relations (\ref{interg}) in particular are of great significance as we shall see below, because they facilitate the study of the exact symmetry of the open spin chain.

Before we continue with the investigation of the symmetry it will be instructive to write explicitly the following coproducts, which are valid for the $gl(n)$ case (see also (\ref{coco2}), (\ref{co}), (\ref{co'}), (\ref{cop2}), (\ref{cop22})) \be (\pi_{\lambda} \otimes \mbox{id}^{\otimes N})\Delta^{(N+1)}(\tilde {\cal Q}_{ab}^{(0)}) &=& \pi_{\lambda}(\tilde {\cal Q}_{ab}^{(0)})
\otimes {\mathbb I} +{\mathbb I} \otimes  \tilde   Q_{ab}^{(0)} \non\\ (\pi_{\lambda} \otimes \mbox{id}^{\otimes N})\Delta^{(N+1)}(\tilde
{\cal Q}_{ab}^{(1)}) &=& \pi_{\lambda}(\tilde {\cal Q}_{ab}^{(1)}) \otimes {\mathbb I} +{\mathbb I} \otimes \tilde  Q_{ab}^{(1)}\non\\
&-&{ (k^*)^{-1}_{cd} \over 2} \Big (\pi_{\lambda}(\tilde  {\cal Q}_{ac}^{ (0)}) \otimes \tilde  Q_{db}^{(0)-} -\pi_{\lambda}( \tilde {\cal Q}_{db}^{(0)}) \otimes \tilde Q_{ac}^{(0)-} \Big )  \non\\ (\pi_{\lambda} \otimes \mbox{id}^{\otimes
N})\Delta^{'(N+1)}(\tilde {\cal Q}_{ab}^{(1)}) &=&
\pi_{\lambda}(\tilde {\cal Q}_{ab}^{(1)}) \otimes {\mathbb I} +{\mathbb I} \otimes \tilde Q_{ab}^{(1)} \non\\&+&{ (k^*)^{-1}_{cd}
\over 2} \Big (\pi_{\lambda}(\tilde  {\cal Q}_{ac}^{(0)-}) \otimes \tilde Q_{db}^{(0)} - \pi_{\lambda}(\tilde {\cal Q}_{db}^{(0)-}) \otimes \tilde Q_{ac}^{(0)} \Big ), \label{cog} \ee where  \be \tilde
Q_{ab}^{(0)-} =\Delta^{(N)}(\tilde {\cal Q}_{ab}^{(0)-}). \ee The crucial point is that equations (\ref{interg}) bear algebraic relations between the entries of the operator ${\cal T}$ and the boundary non-local charges (\ref{tilde}). For simplicity we use the 
XXX ($gl(2)$) model to exhibit the symmetry of the open transfer matrix (\ref{transfer}), although the following procedure may be easily generalized for the $gl(n)$ case. Note that the $gl(2)$  case is quite special, because the twisted Yangian coincides essentially with the boundary Yangian, recall that 
$R_{12}(\lambda) =\bar R_{12}(\lambda)$ (and ${\cal P}=\check {\cal P}$), nevertheless it provides an illuminating paradigm. Let \be {\cal T}_{0}(\lambda) =
\left(
\begin{array}{cc}
{\cal A}_{1} &  {\cal B}\\
{\cal C} &  {\cal A}_{2}   \\
\end{array} \right) ~~\mbox{and} ~~t(\lambda) ={\cal A}_{1}+{\cal A}_{2} 
\label{tr2} \ee then from the commutation relations (\ref{interg}) and also (\ref{gene}), (\ref{eval}), (\ref{cog}) we obtain \be &&\Big [ \tilde Q_{aa}^{(0)},\ {\cal A}_{1}\Big ]= 2i\kappa({\cal B}-{\cal C}), ~~
\Big [\tilde Q_{aa}^{(0)},\ {\cal A}_{2}\Big ]= -2i\kappa({\cal B}-{\cal C}), ~~
a \in \{1,2\} \non\\
&& \Big [\tilde Q_{ab}^{(0)},\ {\cal A}_{1} \Big ] = \Big [\tilde Q_{ab}^{(0)},\
{\cal A}_{2} \Big ] = \Big [ \tilde Q_{ab}^{(0)},\ {\cal B}
\Big ]= \Big [\tilde Q_{ab}^{(0)},\ {\cal C} \Big ]=0, ~~a \neq b \non\\
&& \Big [\tilde Q_{aa}^{(0)},\ {\cal B} \Big ] = 2i\kappa({\cal A}_{1}-{\cal A}_{2}) +2i{\cal B}, ~~\Big [\tilde Q_{aa}^{(0)},\ {\cal C} \Big ] = -2i\kappa({\cal A}_{1}-{\cal A}_{2}) -2i{\cal C}, \label{com20} \ee and it follows that \be \Big [ t(\lambda),\
\tilde Q_{ab}^{(0)} \Big ]= 0, ~~a,b \in \{1, 2\}. \label{com2a} \ee From the intertwining relations (\ref{interg}) and with the help of (\ref{com20}) it also follows that \be && \Big [ t(\lambda),\ \tilde Q_{11}^{(1)} \Big ] = -\Big [ t(\lambda),\
\tilde Q_{22}^{(1)} \Big ] = \Big [ t(\lambda),\  \kappa ( \tilde Q_{12}^{(1)} +\tilde Q_{21}^{(1)})\Big ]= 4\kappa i (\lambda +i) ({\cal B} -{\cal C}), \label{com2} \ee and consequently \be  \Big
[ t(\lambda),\  \tilde Q_{11}^{(1)} +\tilde Q_{22}^{(1)}\Big ]=\Big [ t(\lambda),\  \kappa (\tilde Q_{12}^{(1)} +\tilde Q_{21}^{(1)} )+(-)^{a} \tilde Q_{aa}^{(1)} \Big ]=0. \label{com2b} \ee It should be pointed out that the combinations of non--local charges appearing in (\ref{com2b}) are expressed solely in terms
of $\tilde Q^{(0)}_{ab}$'s, which means that the  only conserved charges entailed so far are the $\tilde Q^{(0)}_{ab}$'s.  In fact the first combination is trivial, because $\tilde  Q^{(1)}_{11} +\tilde  Q^{(1)}_{22} \propto {\mathbb I}$, and it is also expected from the commutation relation (\ref{com}) as $\lambda'
\to \infty$. The existence of higher non-trivial conserved charges is an intriguing question that will be examined in detail elsewhere. It is clear that generalized commutation relations between the generators $\tilde Q_{ab}^{(p)}$, $a,\ b \in \{1,\ldots,n\}$ and the entries of the $gl(n)$ ${\cal T}$ matrix
may be now deduced in an analogous, although technically more complicated way.

It is finally worth emphasizing that the general intertwining relations (\ref{interg}) and the discovered symmetry (\ref{com2a}), (\ref{com2b}) are independent of the choice of representation on the quantum spaces, and 
therefore they are universal results. In the special case where the quantum spaces are mapped via the evaluation representation (\ref{eval}), and ${\cal L} \to R$ the relations (\ref{interg}), (\ref{cog}), (\ref{com2a}), (\ref{com2b}) are of course still valid, but with $\mbox{id}^{\otimes N} \to \pi_{0}^{\otimes N} $.

\section{Discussion}

Let us briefly review the main results of this investigation. The main objective of this work was the study of the remaining symmetries of rational integrable spin chains ($gl(n)$) once non-diagonal integrable boundaries are implemented. We considered two types of boundary conditions known as soliton-preserving and
soliton non-preserving. For both types of boundaries non-local charges (\ref{tilde}) were derived explicitly by means of the study of the asymptotic behaviour of the coproduct type solution of the reflection equation ${\cal T}$. The non-local charges (\ref{tilde}) were simply coproducts of generators of the `boundary' or twisted Yangian (\ref{gene}) depending on the choice of boundary conditions (\ref{re}). Furthermore, by using the intertwining relations (\ref{inter}), (\ref{interg}) we were able to derive the symmetry of the open spin chain (\ref{com2a}). Relations of the form (\ref{inter}) provide  also an alternative way of finding solutions of the reflection equation  (\ref{re}) (see e.g. \cite{dema, dema2}), although there exist other effective algebraic techniques allowing the solution of the reflection equation (see e.g. \cite{doma}).

It should be finally emphasized that $R$ matrices associated to e.g. $o(n)$, $sp(n)$ algebras enjoy crossing symmetry i.e. $R_{12}(\lambda) =\bar
R_{12}(\lambda)$ (\ref{rbar}), and therefore in this case the boundary Yangian coincides with the twisted Yangian (see also \cite{our2, arn}).

\textbf{Acknowledgements:} I am grateful to D. Arnaudon, J. Avan, N. Crampe, L. Frappat and E. Ragoucy for useful discussions on reflection algebras and twisted Yangians. This work is supported by the TMR Network `EUCLID. Integrable models and applications: from strings to condensed matter', contract number HPRN-CT-2002-00325.

\end{document}